\documentclass[aps,pre,twocolumn,showpacs]{revtex4}
\usepackage{amssymb}
\usepackage{amsmath}
\usepackage{epsfig}

\begin{document}

\title{Non-Markovian dynamics of the sine-Gordon solitons}

\author{Lasha Tkeshelashvili}
\affiliation {
Andronikashvili Institute of Physics, Tbilisi State University, 
Tamarashvili 6, 0177 Tbilisi, Georgia}

\begin{abstract}
The sine-Gordon equation exhibits a gap in its linear spectrum. That gives rise to memory, or non-Markovian, effects in the soliton formation processes. The generalized variational approach is suggested to derive the model equation that governs the solitary wave evolution. The detailed analytical and numerical studies show that the soliton relaxation dynamics exhibits the main specific features of quantum emitters decay processes in photonic band gap materials. In particular,
the non-Markovian effects lead to the extremely long-lived oscillations of the sine-Gordon solitons. That results in 
the bound state of the soliton with the radiated linear waves.

\end{abstract}


\maketitle
\section{Introduction}

Nanostructured periodic systems that exhibit a gap in their linear spectrum give rise to a number of unique light-matter interaction processes \cite{busch2007}. Perhaps, the most peculiar example is the strongly modified 
decay dynamics of quantum emitters embedded in such media \cite{lambropoulos2000}. In particular, for the transition frequencies near the band edges pronounced memory, or non-Markovian, effects take place \cite{john1994}. However, similar behavior is expected to be found for other excited systems as well that are coupled to a continuum of states with the band gaps. 

Here, the sine-Gordon equation is considered, which represents an universal model for studies of nonlinear phenomena in many areas of physics \cite{lamb1980}. In dimensionless form it can be written as
\begin{equation}
u_{tt} - u_{xx} + \sin (u) = 0 \, ,
\label{sg} 
\end{equation}
where the subscripts $t$ and $x$ stand for the partial derivatives with respect of time and space variables, respectively.
The fundamental solitary wave solutions of that equation, often called kinks, are given by 
\begin{equation}
u(x,t) = 4 \, \text{atan} \left[ \exp\left( \pm \frac{x - v t}{\sqrt{1 - v^2}} \right) \right] \, .
\label{kink}
\end{equation}
The plus sign corresponds to the kink, and the minus sign to the anti-kink solutions, respectively. These are exponentially localized solitary waves moving with $v$ group velocity. It should be stressed that any higher-order soliton solution of the sine-Gordon model consists of the certain number of fundamental solitons. For example, so-called breather solution represents a bound state of a kink and anti-kink pair \cite{lamb1980}. Therefore, this letter concerns the relaxation dynamics of the fundamental solitons.

For the present discussion it is important that, for small amplitude waves $\sin (u) \approx u$, and therefore, the sine-Gordon model reduces to the linear Klein-Gordon equation which exhibits a band gap \cite{leon2003}. 

One of the fundamental question is to how a given initial condition evolves to the soliton solutions. The sine-Gordon equation is an integrable model and the problem can be solved by means of the inverse scattering method \cite{lamb1980}. The general analysis shows that any initial condition asymptotically relaxes to a certain number of solitons and dispersive radiation. However, in many interesting cases, the formal evaluation of the solution meets technical difficulties, and therefore, is less useful.

In contrast, the variational approach gives simple and explicit, although approximate, expressions for the solitary wave parameters \cite{anderson2001}. However, the original version of the method fails to account for changes caused by the dispersion of the radiation \cite{anderson1983}. That problem was addressed in Ref.~\cite{kath1995}, where the generalized variational ansatz was suggested for studies of the nonlinear Schr\"odinger soliton formation processes. Later, the similar analysis was carried out for the sine-Gordon equation as well \cite{smyth1999}. In should be noted that, as compared to the nonlinear Schr\"odinger equation, worse agreement between the approximate theory and the exact numerical results was achieved. Nevertheless, it was found that, an initially deformed soliton experiences damped oscillations while shedding dispersive radiation. Furthermore, far from the soliton core, the radiated waves obey the linear Klein-Gordon equation provided that the initial deformation is not too strong. 

In what follows, it is demonstrated that the non-Markovian effects, caused by the gap of the linearized spectrum, play a crucial role in the sine-Gordon soliton formation processes. In particular, while relaxing to the steady state, 
initially deformed solitons form a bound state with the radiated linear waves. The similar behavior is known for the decay dynamics of quantum emitters embedded in nanostructures \cite{lambropoulos2000}.

However, first let us introduce a simple model which, as will become apparent below, accounts even quantitatively for the non-Markovian relaxation effects.

\section{The model system}

The suggested model system consists of a harmonic oscillator coupled to a semi-infinite flexible string \cite{beck1960}. The string is assumed to be embedded in an elastic matrix, and the reference frame is chosen such that its rest position  coincides with the positive $x$-axis. Then, the transverse displacement of the string $U(x,t)$ obeys the linear Klein-Gordon equation \cite{morse1953}:
\begin{equation}
U_{tt} - U_{xx} + \omega_{\mathrm{g}}^2 U = 0 \, ,
\label{kg} 
\end{equation}
here $\omega_{\mathrm{g}}$ is determined by the elastic matrix. The ansatz $U(x,t) \sim \exp[-i(\omega t - k x)]$ gives the dispersion relation for the Klein-Gordon model $\omega^2 = \omega_{\mathrm{g}}^2 +k^2$. Therefore, $\omega_{\mathrm{g}}$ fixes the band gap size.

The harmonic oscillator is located at the origin and obeys \cite{beck1960}:
\begin{equation}
\left[ \frac{\mathrm{d}^2}{\mathrm{d} t^2} + \omega_0^2 \right] U(0,t) = 2\gamma U_x(0,t) \, ,
\label{osc} 
\end{equation}
where $\omega_0$ is the oscillator eigenfrequency, and $\gamma$ is related to the tension of the string.

Let us assume that the string is at rest initially. The solution of Eq.~\eqref{kg} for $U(0,t) = f(t)$ boundary condition is well known \cite{sutmann1998}. Using the expressions derived in Ref.~\cite{sutmann1998} we can write
\begin{equation}
U_x(0,t) = - f^{\prime}(t) - \omega_{\mathrm{g}}\int^t_0\frac{J_1(\omega_{\mathrm{g}}\tau)}{\tau}f(t-\tau) \,\mathrm{d} \tau \, ,
\label{Ux} 
\end{equation}
where the prime denotes the time derivative of the corresponding function, and $J_1(t)$ is the first order Bessel function of 
the first kind. In combination with Eq.~\eqref{osc} that gives 
\begin{align}
f^{\prime\prime}(t) + 2\gamma f^{\prime}(t) & + \omega_0^2 f(t) = \nonumber \\
&-2\gamma \, \omega_{\mathrm{g}}\int^t_0\frac{J_1(\omega_{\mathrm{g}}\tau)}{\tau}f(t-\tau) \,\mathrm{d} \tau \, .
\label{model} 
\end{align}
This is a linear integro-differential equation which can be readily solved, for example, by means of the Laplace transform method \cite{morse1953}:
\begin{align}
f(t) &=  \left[ f(0)\frac{\mathrm{d}}{\mathrm{d} t} + 2\gamma f(0) + f^\prime(0) \right] \nonumber \\
&\times\left[ F(t) - \omega_{\mathrm{g}}\int_0^t F\left(\sqrt{t^2 - \tau^2}\right) 
J_1(\omega_{\mathrm{g}} \tau) \, \mathrm{d} \tau \right] \, ,
\label{ft}
\end{align}
where
$$
F(t) = \frac{\exp(r_1 t) - \exp(r_2 t)}{r_1-r_2} \, ,
$$
provided that $r_1\neq r_2$, and
$$
r_{1,2} = -\gamma \pm \sqrt{\gamma^2 + \omega_{\mathrm{g}}^2 - \omega_0^2} \, .
$$
In addition, for $\omega_0^2 = \omega_{\mathrm{g}}^2 +\gamma^2$, i.e. $r_1 = r_2$, we have
$$
F(t) = t\exp(-\gamma t) \, .
$$

The non-Markovian effects in the presented model are introduced through the right-hand side of Eq.~\eqref{model}. For $\omega_{\mathrm{g}} = 0$ that term vanishes, and the governing equation describes a damped oscillator dynamics \cite{landau1979}. In particular, for $\gamma \ll \omega_0$, the system oscillates with the exponentially damped amplitude. For $\gamma \lesssim \omega_0$, however, the damping is so strong that the dynamics is essentially aperiodic. Finally, for $\gamma > \omega_0$, the system experiences exponential relaxation to the equilibrium state with no oscillations. Thus, when the gap is absent, the energy transfer from the oscillator to the string obeys the exponential law.

\begin{figure}[t]
{\epsfig{file=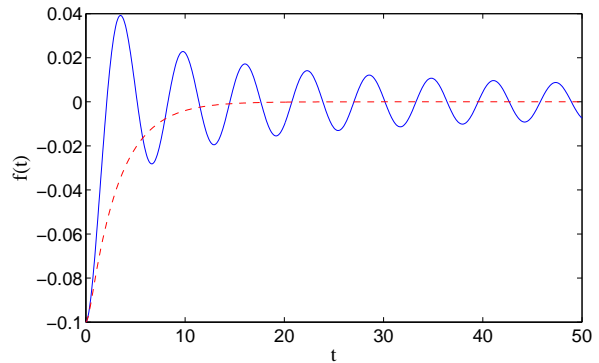,width=8.5cm}} 
\caption{(Color online) The oscillator dynamics for $\omega_{\mathrm{g}} = 0$ (dashed line) and $\omega_{\mathrm{g}} = 1$ (solid line). In both cases $\omega_0 =1.1$, $\gamma = 2$, $f(0) = -0.1$, and $f^\prime(0) = 0$. Further details are discussed in the text. The plotted variables are dimensionless.} 
\label{fig1}
\end{figure}

\begin{figure}[t]
{\epsfig{file=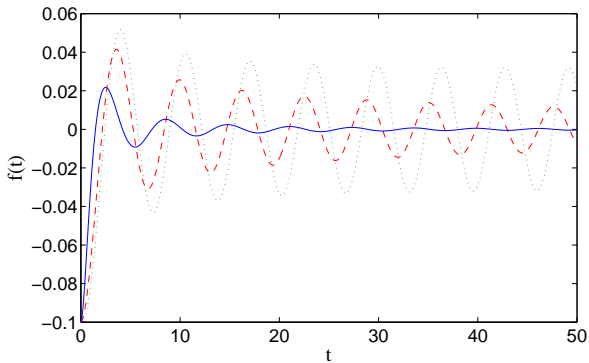,width=8.5cm}} 
\caption{(Color online) The oscillator dynamics for $\omega_0 = 0.5$ (dotted line), $\omega_0 = 1$ (dashed line),  and $\omega_0 = 2$ (solid line). In all cases $\omega_{\mathrm{g}} =1$, $\gamma = 2$, $f(0) = -0.1$, and $f^\prime(0) = 0$. The plotted variables are dimensionless.} 
\label{fig2}
\end{figure}

However, as demonstrated in Fig.~\ref{fig1}, the situation might be completely different for $\omega_{\mathrm{g}} \neq 0$ due to the memory effects. Note that, $\gamma >\omega_0$ in the considered example. In addition, is should be stressed that the oscillator eigenfrequency is outside the gap, but close to the band edge. The system dynamics exhibits the main specific features previously found for quantum emitters decay in photonic band gap materials. First of all, as compared to the free space, the spontaneous decay rate of an excited atom is enhanced for transition frequencies close to the band edge \cite{lambropoulos2000}. For comparison, Fig.~\ref{fig1} shows the relaxation of the oscillator for $\omega_{\mathrm{g}} = 0$. In the presence of gap, for the relatively short time period $t \sim 1$, the system damping rate significantly exceeds that of the gapless case. Second, for the quantum emitters, the photon-atom bound state is present even when the transition frequency is outside the gap \cite{john1994}. The analogous phenomenon can be identified in Fig.~\ref{fig1} as well. Indeed, the oscillations of the system persist even for $t \gg 1$, and the oscillator forms a bound state with the radiated waves. That can be shown by considering the asymptotic behavior of the considered system. In particular, taking the limit $t \rightarrow \infty$ and assuming $f(t) \sim \cos(\omega t)$, let us rewrite Eq.~\eqref{model} as follows
\begin{align}
(\omega_0^2  - \omega^2) & \cos(\omega t) - 2\gamma \omega \sin(\omega t) = \nonumber \\
&-2\gamma \, \omega_{\mathrm{g}}\int^{\infty}_0 \frac{J_1(\omega_{\mathrm{g}}\tau)}{\tau} \cos(\omega t - \omega \tau) \, \mathrm{d} \tau \, .
\label{station} 
\end{align}
This expression determines the allowed values for $\omega$ in the stationary regime. In particular, we have
\begin{equation}
\omega^2 = \omega_0^2 - 2 \gamma^2 \pm 2 \gamma \sqrt{\gamma^2 + \omega_{\mathrm{g}}^2 - \omega_0^2} \, ,
\label{statfreq}
\end{equation}
provided that $\omega$ is real and $\omega \le \omega_{\mathrm{g}}$. Therefore, only those Fourier components of $f(t)$ remain at $t \gg 1$, which correspond to the frequencies given by Eq.~\eqref{statfreq}.
In the considered case Eq.~\eqref{statfreq} allows only one frequency $\omega \approx 0.999$. Therefore, $\omega < \omega_{\mathrm{g}}$ holds, and  the system asymptotically tends to oscillate with the period $T = 2\pi/\omega \approx 6.293$. The bound state is more pronounced for $\omega_0 < \omega_{\mathrm{g}}$, and disappears deeper in the band as shown in Fig.~\ref{fig2}.

Now, let us show that the presented model system describes quantitatively the soliton formation processes for the sine-Gordon equation. 

\section{The variational approach}

Following Ref.~\cite{smyth1999}, let us consider the evolution of the initially deformed anti-kink solution in the reference frame moving with $v$ group velocity 
\begin{equation}
u(x,t) = 4 \, \text{atan} \left[ \exp\left( - \frac{x}{w} \right) \right] \, .
\label{deformed}
\end{equation}
Here, $w = w(t)$ represents the width of the pulse. Note that, the width of the exact anti-kink solution $w_{\mathrm{s}} = 1$ in the chosen reference frame. Moreover, $u \rightarrow 0$ for $x \gg 1$, and $u \rightarrow 2\pi$ for $x \ll -1$.

Let us assume that the initial deformation of the soliton is small $\delta w = (w - w_{\mathrm{s}}) \ll 1$, so that the generated radiation has small amplitude too. Then, the action functional for the problem under consideration reads
\begin{equation}
J = \int_{-\infty}^{+\infty}  L_{\mathrm{nl}} \, \mathrm{d} t + \int_{-\infty}^{+\infty} L_{\mathrm{lin}} \, \mathrm{d} t \, .
\label{action}
\end{equation}
The Lagrange function of the deformed soliton is \cite{smyth1999}:
\begin{equation}
L_{\mathrm{nl}} = \int^{+\infty}_{-\infty} \mathcal{L}_{\mathrm{sg}} \, \mathrm{d} x \, ,
\end{equation}
with the Lagrangian density for the sine-Gordon model
\begin{equation}
\mathcal{L}_{\mathrm{sg}} = \frac{1}{2} u_t^2 - \frac{1}{2}u_x^2 - (1 - \cos u ) \, .
\label{sglagrangian}
\end{equation}

Now, according to the variational approach \cite{anderson2001}, let us insert Eq.~\eqref{deformed} in the Lagrange function for the deformed soliton, to obtain \cite{smyth1999}:
\begin{equation}
L_{\mathrm{nl}} = \frac{\pi^2}{3}\frac{(w^\prime)^2}{w} - \frac{4}{w} - 4w \, .
\label{solvar}
\end{equation}

The linear waves are radiated due to the nonlinear processes in the core region of the soliton 
$ x \in [-\varepsilon, +\varepsilon] $. Therefore, the linear waves are generated at $x = \pm \varepsilon$ and the corresponding  Lagrange function reads
\begin{equation}
L_{\mathrm{lin}} = \int^{+\infty}_{+\varepsilon} \mathcal{L}_{\mathrm{kg}} \, \mathrm{d} x + 
\int^{-\varepsilon}_{-\infty} \mathcal{L}_{\mathrm{kg}} \, \mathrm{d} x \, ,
\end{equation}
where $\mathcal{L}_{\mathrm{kg}}$ is the linearized Lagrangian density of the sine-Gordon model
\begin{equation}
\mathcal{L}_{\mathrm{kg}} = \frac{1}{2} U_t^2 - \frac{1}{2}U_x^2 - \frac{1}{2}U^2 \, .
\end{equation}
That corresponds to the Klein-Gordon equation
\begin{equation}
U_{tt} - U_{xx} + U = 0 \, ,
\label{kggapone}
\end{equation}
with the forbidden gap size $\omega_{\mathrm{g}} = 1$. Thus, assuming that the linear waves obey Eq.~\eqref{kggapone}, for the variation of the action functional we obtain \cite{gelfand1963}:
\begin{align}
\delta J &= \int^{+\infty}_{-\infty} \left[ \frac{\pi^2}{3} \left( \frac{w^\prime}{w} \right)^2 - \frac{2\pi^2}{3} \frac{w^{\prime\prime}}{w} + \frac{4}{w^2} - 4   \right] \delta w \, \mathrm{d} t \nonumber \\
         &+\int^{+\infty}_{-\infty} U_x(+\varepsilon,t) \delta U(+\varepsilon,t) \, \mathrm{d} t \nonumber \\
         &-\int^{+\infty}_{-\infty} U_x(-\varepsilon,t) \delta U(-\varepsilon,t) \, \mathrm{d} t \, .
\label{actionvariation}
\end{align}
The first term on the right-hand side of this expression corresponds to the soliton part \cite{smyth1999}, while the second and third terms result from the variation of $U$ at the boundaries of the soliton core \cite{gelfand1963}.

It should be stressed again that, the linear waves are not generated if $w = w_{\mathrm{s}}$. Therefore, at the soliton core boundaries, $U$ depends on time through $w(t)$. Then, $U(\pm \varepsilon, w)$ can be expanded in a power series of $\delta w$ 
$$
U(\pm \varepsilon, w) = \pm \alpha \, \delta w + \cdots \, , 
$$
where we take into account that $U(\pm \varepsilon, w_{\mathrm{s}}) = 0$, and $ \alpha \equiv d U(\varepsilon, w_{\mathrm{s}}) / d  w$. 
Thus, $U(\pm \varepsilon, w) = \delta U(\pm \varepsilon, w)$, and
\begin{equation}
\delta U(\pm \varepsilon, w) = \pm \alpha \, \delta w \, ,
\label{bogol}
\end{equation}
in the linear approximation. In addition, $u_x(\varepsilon,t) = u_x(-\varepsilon,t)$ as can be seen from Eq.~\eqref{deformed}. The same must hold for $U_x$ as well, and so
\begin{equation}
U_x(\varepsilon,t) = U_x(-\varepsilon,t) \, .
\label{ucapxpm}
\end{equation}

The least action principle $\delta J = 0$, in combination with Eqs.~\eqref{bogol} and \eqref{ucapxpm}, leads to the following governing equation for $w(t)$
\begin{equation}
\frac{2\pi^2}{3} \frac{w^{\prime\prime}}{w} - \frac{\pi^2}{3} \left( \frac{w^\prime}{w} \right)^2 - \frac{4}{w^2} + 4 = 2\alpha \, U_x(\varepsilon, t) \, .
\label{nlvarforw}
\end{equation}

It is clear that $\varepsilon \sim w_{\mathrm{s}}$. Nevertheless, the variational method treats the deformed soliton as a point particle \cite{anderson1983}. For that reason, the actual value for $\varepsilon$ is not needed for our purposes. 
Therefore, for the sake of convenience, formally we can choose $\varepsilon = 0$.  Furthermore, note that $U(x,t)$ obeys Eq.~\eqref{kggapone}, and so, $U_x(\varepsilon = 0,t)$ is given by Eq.~\eqref{Ux} with $\omega_{\mathrm{g}} = 1$ and $f(t) = \alpha \, \delta w(t)$. Finally, since $w = 1 + \delta w$, the left-hand side of Eq.~\eqref{nlvarforw} can be linearized with respect of $\delta w$ to give
\begin{align}
\delta w^{\prime\prime}(t)  + 2\gamma \delta w^{\prime}(t) & + \omega_0^2 \delta w(t) = \nonumber \\
& -2\gamma \int^t_0\frac{J_1(\tau)}{\tau}\delta w(t-\tau)\, \mathrm{d} \tau \, ,
\label{finaldeltaw} 
\end{align}
with $\omega_0 = 2\sqrt{3}/\pi \approx 1.1$, and $\gamma = 1.5\alpha^2/\pi^2$.

\begin{figure}[t]
{\epsfig{file=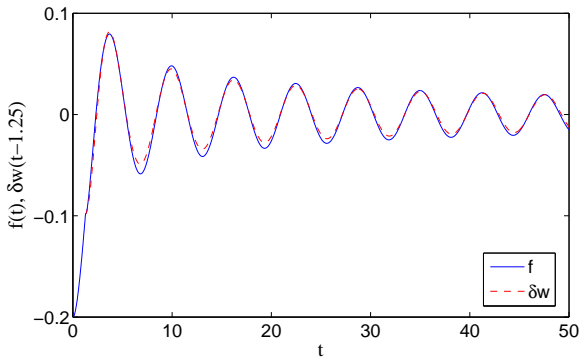,width=8.5cm}} 
\caption{(Color online) The sine-Gordon soliton width variation. The solid line shows $f(t)$, the numerical solution of Eq.~\eqref{model}. $f(0) = -0.2$ and $f^\prime(0) = 0$. $\omega_{\mathrm{g}} = 1$, $\omega_0 = 1.1$, and $\gamma  = 2$. The dashed line gives the phase corrected $\delta w (t -\phi_0)$ determined from the numerical solution of Eq.~\eqref{sg}. $\delta w(0) = -0.1$, $\delta w^\prime(0)=0$, and $\phi_0 = 1.25$. $f^\prime(\phi_0) = \delta w^\prime(0)$ is enforced. The plotted variables are dimensionless.} 
\label{fig3}
\end{figure}

Therefore, the deformed soliton dynamics in the sine-Gordon model is governed by Eq.~\eqref{model}, with the solution given by Eq.~\eqref{ft}. In this case, $\gamma$ represents the coupling constant of the linear radiation to the soliton core. However, since the soliton and dispersive radiation are the constituents of the same solution of the sine-Gordon equation, it is expected that the order of magnitude of $\gamma$ is $1$. Moreover, it is clear that, in the absence of the gap the soliton relaxation process would be exponential. That is, according to the discussion given above, $\gamma > \omega_0 \approx 1.1$. In fact, the choice $\gamma = 2$ guarantees very good agreement with the numerical simulations of the sine-Gordon equation.

\begin{figure}[t]
{\epsfig{file=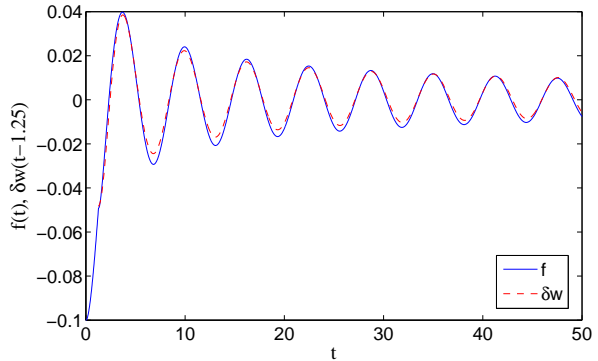,width=8.5cm}} 
\caption{(Color online) The sine-Gordon soliton width variation. The solid line shows $f(t)$, the numerical solution of Eq.~\eqref{model}. $f(0) = -0.1$ and $f^\prime(0) = 0$. $\omega_{\mathrm{g}} = 1$, $\omega_0 = 1.1$, and $\gamma  = 2$. The dashed line gives the phase corrected $\delta w (t -\phi_0)$ determined from the numerical solution of Eq.~\eqref{sg}. $\delta w(0) = -0.05$, $\delta w^\prime(0)=0$, and $\phi_0 = 1.25$. $f^\prime(\phi_0) = \delta w^\prime(0)$ is enforced. The plotted variables are dimensionless.} 
\label{fig4}
\end{figure}

\section{Discussion and conclusions}

There is a subtle point which appears important for comparison of the model calculations with the exact solutions of Eq.~\eqref{sg}. In particular, Eq.~\eqref{Ux} is obtained by solving a boundary value problem with no linear waves at $t=0$. However, the evolution of the deformed solitons for Eq.~\eqref{sg} represents an initial value problem. Therefore, in general, we need to solve a mixed boundary and initial value problem for Eq.~\eqref{kg}. Nevertheless, 
in many interesting cases, the presented boundary value problem for Eq.~\eqref{kg} still can be utilized to obtain the relevant solution for the comparison. For example, consider the initial value problem for Eq.~\eqref{sg} defined by Eq.~\eqref{deformed} with sufficiently small $\delta w(0) \neq 0$ and $\delta w^\prime(0)=0$ \cite{smyth1999}. Note that, the initial deformation extends over the whole space. In the sine-Gordon equation such a perturbation  represents linear radiation everywhere but the soliton core region. In contrast, Eqs.~\eqref{Ux} and \eqref{ft} show that the linear waves are generated for $t>0$. The considered initial condition for Eq.~\eqref{sg} corresponds to a solution of Eq.~\eqref{model} at certain $t = \phi_0$. To show that, the initial phase shift $\phi_0$ between the model calculations and the numerical solution of the sine-Gordon equation must be introduced. Simultaneously, the choice of $f(0)$ must guarantee that $f(\phi_0) = \delta w(0)$. Note that, $\delta w^\prime(0)$ is an arbitrary constant in general. Thus, it is necessary to adjust $f^\prime(\phi_0)$ to $\delta w^\prime(0)$. For this purpose we should solve Eq.~\eqref{model} numerically. 
Then, the solution of the given boundary value problem with the appropriately chosen $\phi_0$, results in the relevant mixed boundary and initial value problem at $t = \phi_0$. Figure~\ref{fig3} shows an example of such calculation. 
Note that, since $f^\prime(\phi_0) = \delta w^\prime(0)$ is enforced, the solution of Eq.~\eqref{model} exhibits a kink at $t = \phi_0$. An excellent quantitative agreement between the numerical solutions of Eqs.~\eqref{sg} and \eqref{model} should be emphasized. The approach developed in Ref.~\cite{smyth1999} gives only qualitative agreement with the exact numerical results.

\begin{figure}[t]
{\epsfig{file=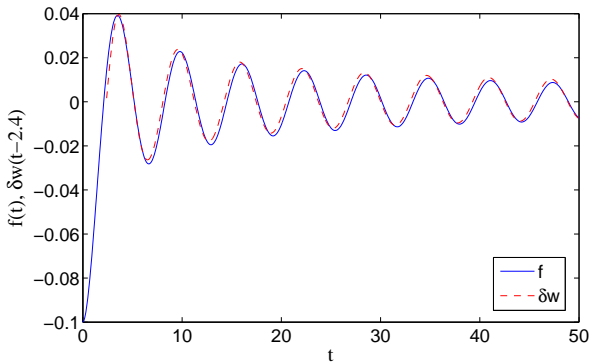,width=8.5cm}} 
\caption{(Color online) The sine-Gordon soliton width variation. The solid line shows $f(t)$ as given by Eq.~\eqref{ft}. 
$f(0) = -0.1$ and $f^\prime(0) = 0$. $\omega_{\mathrm{g}} = 1$, $\omega_0 = 1.1$, and $\gamma  = 2$. The dashed line gives the phase corrected $\delta w (t -\phi_0)$ determined from the numerical solution of Eq.~\eqref{sg}. $\delta w(0) = 0$, $\delta w^\prime(0)=0.05$, and $\phi_0 = 2.4$. The plotted variables are dimensionless.} 
\label{fig5}
\end{figure}

In general, according to the discussion given above, the phase shift $\phi_0$ depends on the initial perturbation. 
However, as demonstrated in Fig.~\ref{fig4}, for the considered initial value problem $\phi_0 = 1.25$ for other values of $\delta w(0)$ as well. That is not surprising since Eq.~\eqref{finaldeltaw} is linear. Nevertheless, for $\delta w(0) = 0$ and $\delta w^\prime(0) \neq 0$ the phase shift $\phi_0$ is different. An example is presented in Fig.~\ref{fig5}. In contrast, it must be stressed that $\gamma = 2$ holds for any initial value problem.

In the present article the evolution of the initially deformed solitons is treated in the reference frame moving with $v$ group velocity. That is possible if the solitons travel with constant velocity while relaxing to the steady state.  
The relativistic invariance of the sine-Gordon model guarantees that this is the case. Indeed, let us consider the deformed soliton with $v=0$. From the symmetry of the stated problem it directly follows that the soliton does not accelerate during the relaxation process. Now, consider the same soliton in the reference frame moving with arbitrary velocity  $-v$. In that moving reference frame the soliton travels with the group velocity $+v$. Since the soliton does not accelerate in the original reference frame, according to the special relativity, it does not accelerate in any other inertial reference frame too. That is, while relaxing, the moving deformed soliton propagates with the constant velocity.

The studied soliton oscillations at $ t \gg 1 $ have different origin as compared to the wobbling kink solutions of the $\phi^4$ and the sine-Gordon models found in Ref.~\cite{segur1983}. Indeed, in the case of $\phi^4$ model, the kink oscillations are due to the linear discrete eigenmode of the system on the static kink background. The sine-Gordon kinks do not possess such eigenstates. The wobbling kink of the sine-Gordon model represents a three-soliton solution, and so, is an intrinsically nonlinear excitation. In contrast, the extremely long-lived oscillations of the sine-Gordon solitons studied here are governed by the linear equation. The deformed solitons asymptotically tend to the stationary oscillatory state which is caused solely by the forbidden band gap. That represents the bound state of the soliton with the radiated linear waves.

Finally, note that, the memory effects can be identified in the nonlinear Schr\"odinger soliton formation processes as well \cite{kath1995}. The linear spectrum of that equation does not possess a forbidden gap. Nevertheless, in certain limiting cases it correctly accounts for the effects associated with band gaps. For instance, that is the case for nonlinear photonic band gap materials \cite{desterke1988, desterke1994}. Furthermore, in the weakly nonlinear regime, the sine-Gordon model reduces to the nonlinear Schr\"odinger equation \cite{oikawa1974}. Therefore, in that limiting case of the sine-Gordon equation, it should not be surprising to find the effects reminiscent of non-Markovian dynamics. 

In conclusion, a generalized variational approach is developed to model the soliton formation processes for the sine-Gordon model. It is shown that, if the initial soliton deformation is not too strong, the system dynamics is governed by the linear integro-differential equation. The detailed analytical and numerical examination of the suggested model indicates that the soliton relaxation dynamics exhibits the main specific features of quantum emitters decay processes in photonic band gap materials. The presented results unveil the significant role of the non-Markovian effects in the sine-Gordon soliton formation dynamics.

\section*{Acknowledgments} 

This work is supported by Georgian National Science Foundation (Grant No.~30/12).

\end{document}